\documentclass[conference]{IEEEtran}
\IEEEoverridecommandlockouts
\usepackage{cite}
\usepackage{amsmath,amssymb,amsfonts}
\usepackage{algorithmic}
\usepackage{graphicx}
\usepackage{textcomp}
\usepackage{xcolor}

\def\BibTeX{{\rm B\kern-.05em{\sc i\kern-.025em b}\kern-.08em
    T\kern-.1667em\lower.7ex\hbox{E}\kern-.125emX}}
\begin{document}

\title{Design Optimization of Permanent-Magnet Based Compact Transport Systems for Laser-Driven Proton Beams
\thanks{The work was supported by Laboratory Directed Research and Development (LDRD) funding from LBNL provided by the Director, by the U.S. Department of Energy Office of Science, Offices of Fusion Energy Sciences and High Energy Physics under Contract No. DE-AC02-05CH11231, and by LaserNetUS (https://www.lasernetus.org/). J.~T.~De Chant was supported by the US Department of Energy, Office of Science, High Energy Physics under Cooperative Agreement award number DE-SC0018362.

*Also at Michigan State University
}
}

\author{\IEEEauthorblockN{1\textsuperscript{st} Jared T. De Chant*}
\IEEEauthorblockA{
\textit{Lawrence Berkeley National Laboratory}\\
Berkeley, CA USA \\
jtdechant@lbl.gov}
\and
\IEEEauthorblockN{2\textsuperscript{nd} Kei Nakamura}
\IEEEauthorblockA{
\textit{Lawrence Berkeley National Laboratory}\\
Berkeley, CA USA \\
knakamura@lbl.gov}
\and
\IEEEauthorblockN{3\textsuperscript{rd} Qing Ji}
\IEEEauthorblockA{
\textit{Lawrence Berkeley National Laboratory}\\
Berkeley, CA USA \\
qji@lbl.gov}
\and
\IEEEauthorblockN{4\textsuperscript{th} Lieselotte Obst-Huebl}
\IEEEauthorblockA{
\textit{Lawrence Berkeley National Laboratory}\\
Berkeley, CA USA \\
lobsthuebl@lbl.gov}
\and
\IEEEauthorblockN{6\textsuperscript{th} Samuel K. Barber}
\IEEEauthorblockA{
\textit{Lawrence Berkeley National Laboratory}\\
Berkeley, CA USA \\
sbarber@lbl.gov}
\and
\IEEEauthorblockN{7\textsuperscript{nd} Antoine M. Snijders}
\IEEEauthorblockA{
\textit{Lawrence Berkeley National Laboratory}\\
Berkeley, CA USA \\
amsnijders@lbl.gov}
\and
\IEEEauthorblockN{8\textsuperscript{nd} Thomas Schenkel}
\IEEEauthorblockA{
\textit{Lawrence Berkeley National Laboratory}\\
Berkeley, CA USA \\
t\_schenkel@lbl.gov}
\and
\IEEEauthorblockN{9\textsuperscript{nd} Jeroen van Tilborg}
\IEEEauthorblockA{
\textit{Lawrence Berkeley National Laboratory}\\
Berkeley, CA USA \\
jvantilborg@lbl.gov}
\and
\IEEEauthorblockN{10\textsuperscript{th} Cameron G. R. Geddes}
\IEEEauthorblockA{
\textit{Lawrence Berkeley National Laboratory}\\
Berkeley, CA USA \\
cgrgeddes@lbl.gov}
\and
\IEEEauthorblockN{11\textsuperscript{th} Carl B. Schroeder}
\IEEEauthorblockA{
\textit{Lawrence Berkeley National Laboratory}\\
Berkeley, CA USA \\
cbschroeder@lbl.gov}
\and
\IEEEauthorblockN{12\textsuperscript{th} Eric Esarey}
\IEEEauthorblockA{
\textit{Lawrence Berkeley National Laboratory}\\
Berkeley, CA USA \\
ehesarey@lbl.gov}
}

\maketitle

\begin{abstract}
Laser-driven (LD) ion acceleration has been explored in a newly constructed short focal length beamline at the BELLA petawatt facility (interaction point 2, iP2). For applications utilizing such LD ion beams, a beam transport system is required, which for reasons of compactness be ideally contained within 3 m. The large divergence and energy spread of LD ion beams present a unique challenge to transporting them compared to beams from conventional accelerators. This work
gives an overview of proposed compact transport designs that can satisfy different requirements
depending on the application for the iP2 proton beamline such as radiation biology, material science, and high energy density science. These designs are optimized for different parameters such as energy spread and peak proton density according to an application’s need. The various designs consist solely of permanent magnet elements, which can provide high magnetic field gradients on a small footprint. While the field strengths are fixed, we have shown that the beam size and energy can be tuned effectively by varying the placement of the magnets. The performance of each design was evaluated based on high order particle tracking simulations of typical LD proton beams. A more detailed investigation was carried out for a design to deliver 10~MeV LD accelerated ions for radiation biology applications. With these transport system designs, the iP2 beamline is ready to house various application experiments.

\end{abstract}


\section{Introduction}
Laser-driven (LD) ion acceleration has recently been explored in a newly constructed short-focal length beamline at the BELLA petawatt laser facility, named interaction point 2 (iP2) \cite{Hakimi_2023}, which is available to users through LaserNetUS. In LD ion acceleration, a high intensity laser pulse ($>10^{21}$~W/cm$^2$) 
interacts with a thin solid target, generates a plasma, and accelerates a large number of ions through various mechanisms. This acceleration technique differs from conventional RF cavity driven ion acceleration due to its high ion production ($\sim10^{13}$ ions per shot) and ultra short bunch length ($\sim$ps)~\cite{Macchi_2017}. These properties make LD ion beams attractive
for certain applications, such as high dose rate radiobiology~\cite{Bulanov_2002} and isochoric heating~\cite{Patel_2003}. 

For the new iP2 area at BELLA, a compact proton beam transport system is needed for experiments exploring these applications. As different applications require different beam parameters, it is necessary to explore many different transport configurations to better respond to users' needs. For example, proton therapy experiments require a uniform dose distribution over a relatively large area
on target, while isochoric heating studies would more benefit from maximizing charge intensity on target. 

Transporting LD ion beams has proved difficult due to the large divergence and energy spread of the beam. To accommodate these beams, high field gradient focusing elements have been used in the field, such as pulsed solenoids~\cite{Burris-Mog_2011,Jahn_2018} and active plasma lenses~\cite{Panofsky_1950, vanTilborg_2015, Lindstrom_2018}, and permanent magnet quadrupoles~\cite{Lim_2005, Eichner_2007, Nishiuchi_2010,Bin_2012}.  Due to available real estate surrounding the iP2 target chamber, the total length of transport from the ion source, including beam optics, diagnostics, and ion target, must be confined to 3~m. This constraint further complicates the design of the transport and is part of what motivated this study to focus only on the use of permanent magnet elements in the transport design, due to their compact size, high field strength, and lower cost.

This paper gives an overview of the methodology of designing the transport and highlights a few of the different configurations that were simulated to evaluate performance. This work allows the iP2 beamline to better respond to the needs of users and their various applications. A final configuration of the beam transport designed to deliver 10~MeV LD ion beams is also shown.

\section{Design Methodology}
\subsection{Beam parameters}
In the simulations in the following sections, the properties of the beam coming out of the laser-plasma interaction region was modeled based on previous experiments and other LD ion accelerators. The initial beam was modeled as having a 2D Gaussian spatial distribution with a full-width-half-maximum (FWHM) beam size of 100~$\mu$m and a beam divergence of 450~mrad \cite{Schreiber_2004,Bin_2022}. 
The angular divergence may depend on the particle energy~\cite{Bin_2013} though here it was treated as constant. The energy spectrum was modeled as a Maxwell-Boltzmann distribution with a mean temperature of 7 MeV which is consistent with experimental ion acceleration results in the TNSA regime with on-target laser intensity of $>10^{20}$~W/cm$^2$ \cite{Macchi_2013}, 

\begin{equation}\label{eq:boltz}
    N(E)/\text{MeV}=5 \times 10^{10} e^{-E/(7 \text{ MeV})}.
\end{equation}

 Particles with an initial divergence larger than the angular acceptance of the first magnet are neglected for the rest of the calculations but are used to estimate the collection efficiency. After the initial beam is prepared, the trajectory is simulated through a transport configuration. The performance of each transport design was then evaluated by estimating the peak density, energy acceptance, and beam spot size of the output beam.

\subsection{Permanent magnet optical elements}

This study uses permanent magnet optical elements to focus and deflect the ion beam. Sets of permanent magnets with specific magnetization vectors can be arranged in a Halbach array~\cite{Halbach_1980} to achieve the same field profiles seen in typical electromagnetic optical elements (e.g. dipoles, quadrupoles, and sextupoles) but with higher fields. In general, permanent magnet quadrupoles can achieve gradients of 100s T/m while their normal conducting electro-magnetic counterpart is typically limited to 1 T/m. Superconducting magnets~\cite{Wan_2015} produce a very strong field gradient but run the risk of quenching from being so close to the interaction point.

Permanent magnet quadrupoles (PMQs) are able to achieve this field strength with a relatively large bore (~20 mm) which is important as a large bore leads to a larger admittance. And as the applications for LD accelerated beams generally require a large beam current, maximizing collection efficiency is crucial for the transport design. 

\begin{figure}[htbp]
\centerline{\includegraphics[width = .4\textwidth]{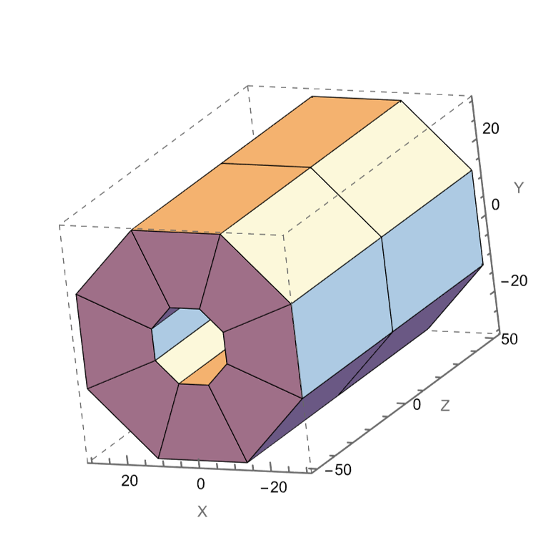}}
\caption{\label{fig:PMQ} Visualization of permanent magnet quadrupole. The parameters are described in Table~\ref{tab:PMQ}.}
\end{figure}

The magneto-static simulation code RADIA~\cite{Chubar_1998} was used to design the PMQs and an example geometry is shown in Fig.~\ref{fig:PMQ}. The magnets were based off an 8-azimuthal-segment Halbach array with 1.29~T of the bulk magnetisation $B_r$. This configuration is readily achievable for commercial vendors, and this bulk magnetisation level is typical for a N40 grade NdFeB magnet, which is also commercially available at relatively low cost. A list of properties of the PMQs used in the design work is shown in Table~\ref{tab:PMQ}. The use of permanent magnet dipoles (PMDs) was also investigated as they are energy dispersive and thus can be used to deflect ions of undesired energy off of the main beam path.  The properties of the PMDs used in this work are shown in Table \ref{tab:PMDs}. 

Permanent magnet optical elements are not easily tunable as their field gradients are fixed after fabrication. However, the properties of the beam can still be tailored to suit the needs of applications. Varying the number of magnets and their relative spacing allows for some flexibility in the beam profile and energy distribution.

\begin{table}[htbp]
\caption{Properties of the permanent magnet quadrupoles}
\label{tab:PMQ}
\begin{center}
\begin{tabular}{lr}
\hline\hline
\multicolumn{2}{c}{\textbf{PMQ parameters}} \\
\hline
Inner radius & 10~mm \\
Outer radius & 30~mm \\
Bulk magnetisation $B_r$ & 1.29~T\\
Effective magnetic field at tip $B_{tip}$ & 1.16~T\\
\hline\hline
\end{tabular}
\end{center}
\end{table}

\begin{table}[htbp]
\caption{Properties of the permanent magnet dipoles}
\label{tab:PMDs}
\begin{center}
\begin{tabular}{lr}
\hline\hline
\multicolumn{2}{c}{\textbf{PMD parameters}} \\
\hline
Half gap & 10~mm \\
Bulk magnetisation $B_r$ & 1.29~T\\
Effective magnetic field $B_{D}$ & 0.89~T\\
\hline\hline
\end{tabular}
\end{center}
\end{table}

\subsection{Simulation}
The design of each transport configuration was done using COSY INFINITY~\cite{COSY_INFINITY} (COSY), an arbitrary order beam dynamics simulation and analysis code. COSY allows for the study of accelerator lattices, spectrographs, beam transports, electron microscopes, and many other devices. With differential algebraic techniques, it calculates Taylor maps, a representation of the motion of charged particle in an arrangement of electromagnetic fields, to arbitrary order. Applying the map, $M$, of some transport lattice to the phase-space vector describing the initial beam distribution, $\vec{z_0}$, produces an estimate of the phase-space vector of the beam after having passed through the transport $\vec{z_f}$. 

\begin{equation}
    \vec{z_f}=M(\vec{z_0}).
\end{equation}

In COSY, each configuration of optical elements was optimized to achieve either point-to-point focusing or collimation at the target location. The COSY output maps were used to evaluate a system's angular acceptance and transmission efficiency. In order to analyze the performance of the transport systems with arbitrary input ion beams, a home-made MATLAB script was written. This script applies the COSY map to the phase-space vectors of a large number of particles to simulate their trajectories and calculate macro-properties of the beam, such as the charge density profile and energy spread. 

Due to the large energy spread and divergence of the initial beam, simulations had to be performed at 3rd order or above to account for spherical and chromatic aberrations inherent to the charged particle optics that would not be included in a linear mapping. 

\section{Results}
\subsection{Comparing doublet, triplet, and quartet configurations}

\begin{figure}[htbp]
\centerline{\includegraphics[width = .5\textwidth]{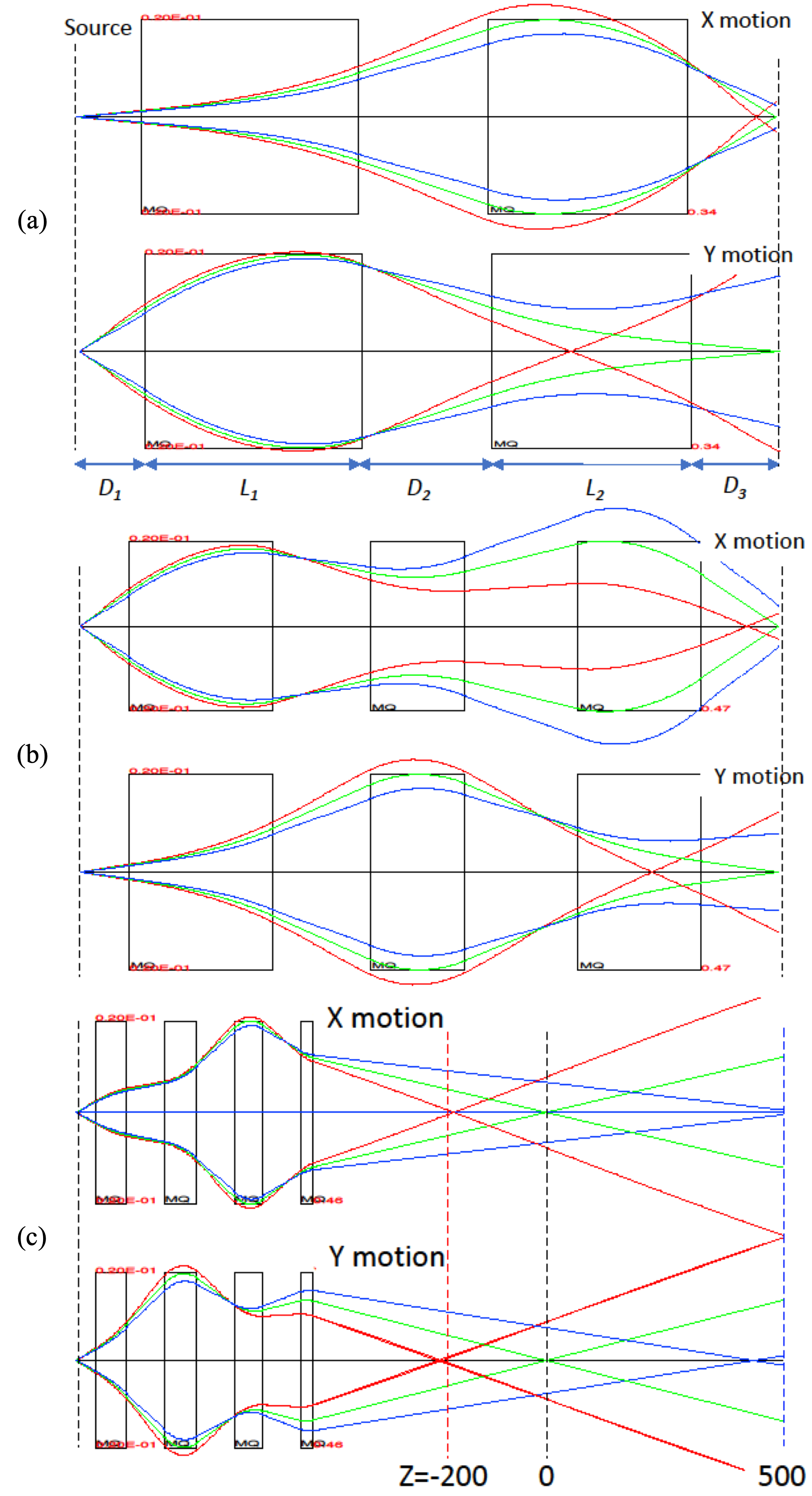}}
\caption{The horizontal and vertical trajectories of particles with the maximum accepted angle in the doublet (a), triplet (b), and quartet (c) configurations. The beam moves from left to right. The different colors show the trajectories of similar particles with energies 20\% above (blue) and 20\% below (red) the nominal beam energy of 30~MeV (green). The main energy of the beam can be tuned by shifting the target location. }
\label{fig:tracking}
\end{figure}

Various configurations of transport elements were investigated to achieve point-to-point focusing of a 30 MeV ion beam. This was done to explore the general costs and benefits of varying the number of magnets. A selection of the designs are shown here, referred to as the doublet, triplet, and quartet corresponding to the number of PMQs present in the configuration. The horizontal and vertical trajectories of particles with the maximum initial angle that is accepted by the transport are shown for each of the configurations in Fig.~\ref{fig:tracking}. The different colors show the trajectories of similar particles with energies 20\% above (blue) and 20\% below (red) the nominal beam energy of 30~MeV (green). Real quadrupoles focus off-energy particles to different focal positions from the nominal focus. This can only be seen in higher order expansions of the quadrupole focusing kick. 

The performance of each configuration was evaluated by estimating the collection efficiency $\eta$, peak proton density $n_{pk}$, momentum acceptance $\delta E$, and FWHM beam size in $x$ ($\sigma_x$) and $y$ ($\sigma_y$). The results for the selection of configurations is shown in Table~\ref{tab:results} to compare.

\begin{table}[htbp]
\caption{Simulation results of collection efficiency $\eta$, peak proton density $n_{pk}$, energy acceptance $\delta E$, and FWHM beam size in $x$ ($\sigma_x$) and $y$ ($\sigma_y$) propagated through the different PMQ configurations investigated in this work.}
\label{tab:results}
\begin{center}
\begin{tabular}{l|ccccc}
\hline\hline
 & $\eta$ (\%) & $n_{pk}$ (10$^9$/mm$^2$) & $\delta E$ (MeV) & $\sigma_{x}$ (mm) & $\sigma_{y}$ (mm) \\
 \hline
Doublet              & 2.2      & 18            & 0.75       & 0.84    & 0.03    \\
Triplet              & 2.5      & 25            & 0.85       & 0.13    & 0.17    \\
Quartet              & 1.3      & 6.6           & 1.2        & 0.34    & 0.35   \\
\hline\hline
\end{tabular}
\end{center}
\end{table}

The simplest configuration is a doublet, or two quadrupole magnets in succession with drift space in between. Two are required at minimum as quadrupole magnets are simultaneously focusing in one plane and defocusing in the other but can be combined together to create overall focusing. As seen in Table \ref{tab:results}, the doublet configuration has a high collection efficiency ($>$2\%) but produces a very asymmetric beam shape on the target. This configuration may be sufficient for applications that mostly require a high beam intensity such as isochoric heating. 

Adding a third magnet improves the collection efficiency, charge density, and symmetry of the beam spot as the extra magnet adds further control over the beam. This configuration may be sufficient for most applications as it performs moderately well in all areas. 

A fourth magnet reduces the overall acceptance of the transport and thus reduces the collection efficiency and charge density. However, the beam spot is nearly perfectly symmetric and the energy acceptance is the highest of the three. This configuration may be more suited for applications such as radiobiology as it would produce a more controlled dose deposition. 

\subsection{Energy tunability}
The main energy of the beam delivered onto target can be tuned in a few different ways. For each of the configurations, the drift lengths between the magnets can be varied so that the nominal beam energy is focused at the target location. Alternatively, the target location can be moved to match the focal position of a particular beam energy as the focal positions of the quadrupoles are energy dependent. This scheme would only work for the quartet configuration, however, as any beam spot asymmetry grows worse the farther away from nominal focus. 


The most effective way to tune and select the beam energy is by passing the beam through a magnetic dipole to disperse the beam depending on the particle energy, similar to a magnetic spectrometer. By adding an aperture at a certain deflection angle, only particles with the desired energy will make it through to the target. By chaining a series of dipoles together in a magnetic chicane or fragment separator type arrangement, the energy can be selected more precisely. Simulations were performed for these configurations but were not included in this presentation.

\subsection{10~MeV Collimator at iP2}
\begin{figure*}[htbp]
\centerline{\includegraphics[width = .8\textwidth]{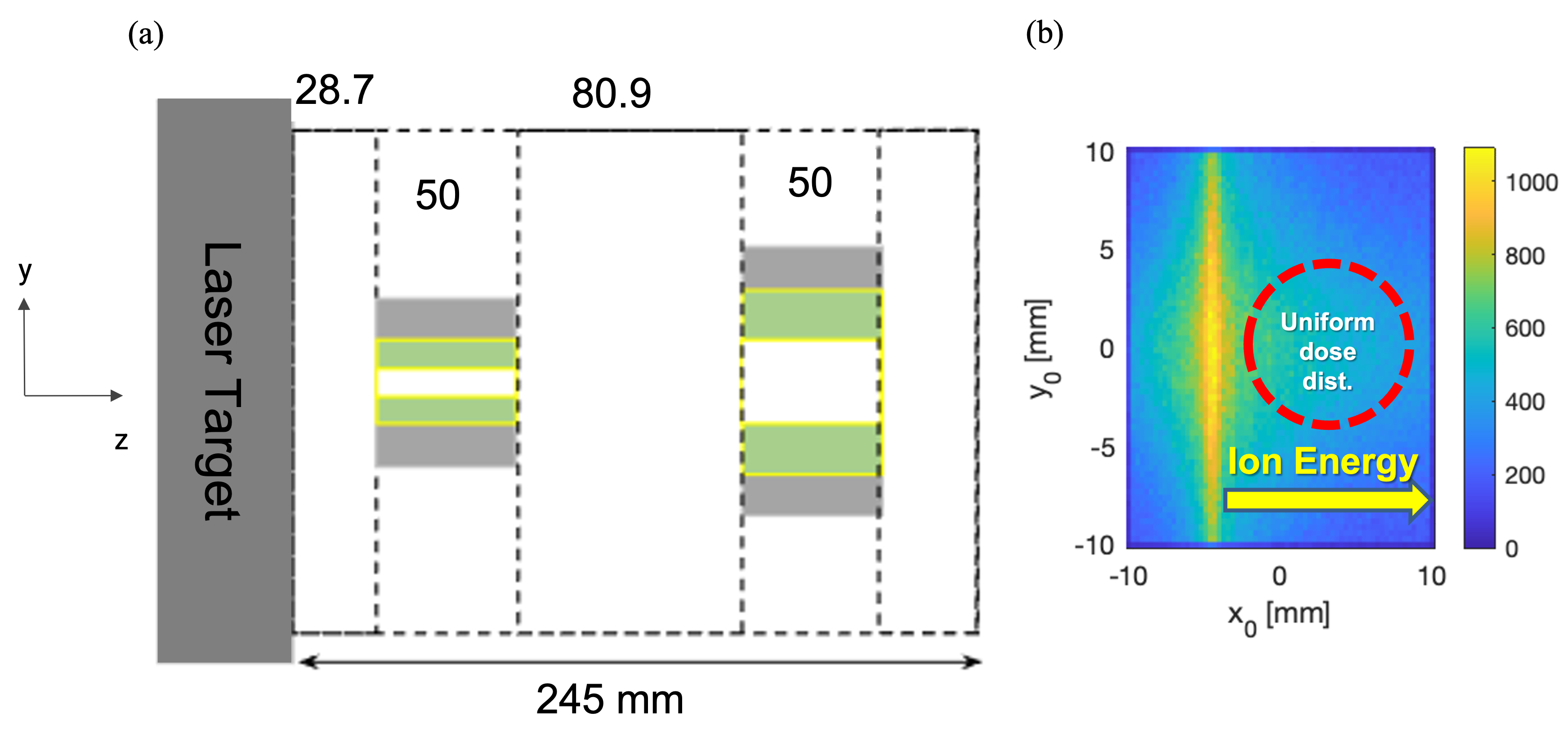}}
\caption{(a) 10 MeV Collimator magnet geometry (dipole not shown). (b) Charge distribution (arbitrary units) at the proton sample location. A region of relatively uniform dose is achieved to the side from where the beam is focused.}
\label{fig:10MeVColl}
\end{figure*}

The first application experiment at iP2 is a radiobiological application utilizing 10~MeV proton beams, which requires a uniform dose distribution over a centimeter level at the sample location. Since oblique laser incidence on target~\cite{Hakimi_2023} is planed, available space is limited further. In order to overcome constraints, a scheme with a compact collimating doublet with a dipole magnet is considered. A collimating doublet instead of a focusing doublet provides large beam size with the minimal real estate. A dipole was used to prevent neutrons and X-Rays from reaching to the samples~\cite{Bin_2022}.
The PMQ geometries and field strengths were also optimized to maximize collection efficiency and are shown in Fig.~\ref{fig:10MeVColl} and Table~\ref{tab:10MeVColl}. 

\begin{table}[htbp]
\caption{Properties of the PMQs for 10 MeV Collimator}
\label{tab:10MeVColl}
\begin{center}
\begin{tabular}{lcc}
\hline\hline
& \textbf{PMQ1} & \textbf{PMQ2} \\
\hline
Inner radius (mm) & 5 & 15\\
Outer radius (mm) & 15 & 45\\
Gradient (T/m) & 250 & 67\\
Effective length (mm) & 50 & 50\\
Integrated Field Strength (T) & 12.3 & 3.35 \\

\hline\hline
\end{tabular}
\end{center}
\end{table}

The dipole was positioned 1.5~m downstream of the second quadrupole to balance the charge density and uniformity. As the configuration consists of a doublet, it produces an asymmetric beam spot. To remedy this, the dipole would be oriented in a way to provide a kick perpendicular to the flat beam to "spread out" the dose distribution. This scheme is illustrated in Fig.~\ref{fig:10MeVColl}. This design can provide a compact and cost-effective beam transport for radiobiological applications.

\section{Conclusion}
This paper explored various compact beam transport designs utilizing permanent magnets. This study has built a strong framework that allows the BELLA Center to swiftly model and adapt the beamline to new experimental requirements and requests from users of the facility. The iP2 beamline is ready to house various application experiments and furthermore, those designs can be used in any other facilities especially where a compact beam transport system is desired.  

\section*{Acknowledgment}
Authors would like to thank Sven Steinke for his contributions to the iP2 beamline.

\bibliographystyle{ieeetr}
\bibliography{citations}

\end{document}